\documentclass{article}
\usepackage{standalone}
\usepackage{amsmath,amssymb}
\usepackage{amsthm}
\usepackage{algorithmic}
\usepackage{subcaption}
\usepackage{algorithm}
\usepackage{arxiv}
\usepackage{hyperref}       
\usepackage{url}            
\usepackage{booktabs}       
\usepackage{amsfonts}       
\usepackage{nicefrac}       
\usepackage{microtype}      
\usepackage{lipsum}		
\usepackage{graphicx}
\usepackage{wasysym}

\newcommand*{\vpointer}{\hbox{\scalebox{1}{\Huge\pointer}}}
\usepackage[
    natbib=true,
    style=numeric,
    sorting=none
]{biblatex}
\hypersetup{
    colorlinks=true,
    linkcolor=blue,
    filecolor=blue,      
    urlcolor=blue
}
\newcommand{\be}{\begin{equation}}
\newcommand{\ee}{\end{equation}}
\newcommand{\bea}{\begin{eqnarray}}
\newcommand{\eea}{\end{eqnarray}}
\newcommand{\bean}{\begin{eqnarray*}}
\newcommand{\eean}{\end{eqnarray*}}

\title{Efficient CPU-Optimized Parameter Estimation for Modeling Fish Schooling Behavior in Large Particle Systems}
\author{Salah Alrabeei\\
Department of Computer Science\\
Western Norway Univ. of Applied Sciences\\
	Bergen, Norway\\
 \texttt{salah.alrabeei@hvl.no}\\
	\And
	\href{https://orcid.org/0000-0003-0646-9830}{\includegraphics[scale=0.06]{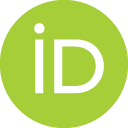}\hspace{1mm}Sam~Subbey}\thanks{Also at Department of Computer Science, Western Norway Univ. of Applied Sciences,	PO Box 7030, 5020 Bergen, Norway}\\
	Inst. of Marine Research\\
	PO Box 1870\\
	5817 Bergen, Norway\\
	\texttt{samuels@imr.no}\\
}

\hypersetup{
pdftitle={A template for the arxiv style},
pdfsubject={q-bio.NC, q-bio.QM},
pdfauthor={David S.~Hippocampus, Elias D.~Striatum},
pdfkeywords={First keyword, Second keyword, More},
}

\begin{document}
\maketitle
\begin{abstract}
The schooling behavior of fish can be studied through simulations involving a large number of interacting particles. In such systems, each individual particle is guided by behavior rules, which include aggregation towards a centroid, collision avoidance, and direction alignment. The movement vector of each particle may be expressed as a linear combination of behaviors, with unknown parameters that define a trade-off among several behavioral constraints. A fitness function for collective schooling behavior encompasses all individual particle parameters.

For a large number of interacting particles in a complex environment, heuristic methods, such as evolutionary algorithms, are used to optimize the fitness function, ensuring that the resulting decision rule preserves collective behavior. However, these algorithms exhibit slow convergence, making them inefficient in terms of CPU time cost.

This paper proposes a CPU-efficient iterative (Cluster, Partition, Refine -- CPR) algorithm for estimating decision rule parameters for a large number of interacting particles. In the first step, we employ the K-Means (unsupervised learning) algorithm to cluster candidate solutions. Then, we partition the search space using Voronoi tessellation over the defined clusters. We assess the quality of each cluster based on the fitness function, with the centroid of their Voronoi cells representing the clusters. Subsequently, we refine the search space by introducing new cells into a number of identified well-fitting Voronoi cells. This process is repeated until convergence.

A comparison of the performance of the CPR algorithm with a standard Genetic Algorithm reveals that the former converges faster than the latter. We also demonstrate that the application of the CPR algorithm results in a schooling behavior consistent with empirical observations.
\end{abstract}
\section{Introduction}
\label{sec:introduction} 
The proposed approach overcomes the slow-convergence weakness of stochastic algorithms due shrinkage of the search space as the algorithm evolves, and hence, the probability of getting the optimal solution increase. 

Simulating complex systems, such as schooling in fish, flocking in birds, and herding or crowding in mammals, provides examples of collective behavior in animals. A simulation framework for understanding such collective behavior has been based on an agent-based modeling approach, which remains an active area of research in computational simulations.
In the Boid (or Relond) model \cite{reynolds1987flocks} -- one of the earliest of such models -- each individual aligns its directional movement with its neighbors, moves away from those neighbors who are too close to avoid collisions, and then flocks together. Later extensions of the Boid model incorporate features such as avoidance of obstacles \cite{erra2004massive,tran2020switching} or predators \cite{barksten2013extending,chang2019investigating}, leadership \cite{hartman2006autonomous,alaliyat2014optimisation}, field preference attraction \cite{chen2006genetic,dmytruk2021safe}. 
The directional heading (moving vector) of each individual boid is determined by a weighted linear combination of the behavior rules, and the model's complexity increases with an increasing number of rules. A challenge in deriving realistic fish schooling behavior is how to choose the weights (coefficients) associated with each rule.

Several stochastic optimization algorithms have been utilized to optimize flocking behaviors \cite{alaliyat2014optimisation,chen2006genetic,alaliyat2019optimal,samarasinghe2022grammar}. In general, Genetic Algorithms (GAs) are one of the most widely used stochastic algorithms \cite{goldberg1989genetic}. GAs apply the principles of natural evolution to the problem of finding an optimal solution \cite{chen2006genetic,goldberg1989genetic,forrest1993genetic}. In general, since the objective function is defined based on the simulation output, the CPU-time cost can be very high when using stochastic algorithms, especially for high-dimensional systems.
Specifically for GAs, the rate of convergence (measured in CPU-time) is strongly determined by the size and dimension of the parameter space. This is because the search space, when using a GA, is static throughout the evolution process. In other words, the probability of obtaining the optimal solution, either at the first generation or the last generation, is identical.

In this paper, we propose a new optimization algorithm to optimize fish schooling behavior rules by finding the optimal weights of the fish moving vector. Our approach is based on an iterative process that utilizes the K-means algorithm \cite{macqueen1967some,sun2008clustering} to generate clusters of candidate solutions from the search space. This is followed by a partitioning of the search space using Voronoi tessellations \cite{boots2009spatial}, where the center of each Voronoi cell is defined by the centroid of the clustered solutions embedded in that cell. We use the principles of natural evolution (selection, mutations, cross-over) to find the optimal solution (Voronoi cell) from the resulting (clustered and partitioned) search space. The proposed approach overcomes the slow-convergence weakness of stochastic algorithms due to the shrinkage of the search space as the algorithm evolves, and hence, the probability of getting the optimal solution increases. This paper aims to demonstrate the efficacy of the proposed algorithm in optimizing fish schooling behavior and its performance efficiency in terms of the rate of convergence when compared to a conventional GA.

This paper is organized as follows: the description of fish schooling model is presented in Section~\ref{sec2}, while the K-Means algorithm and a description of Voroni tessellations are presented in Section~\ref{sec3}. A detailed description of the CPR algorithm, and proposed numerical experiments involving the application of the algorithm are presented in Section~\ref{sec4}. We discuss results from the numerical experiments in Section~\ref{sec5}.
\section{Description of fish schooling model}
\label{sec2}
\begin{figure}[!ht]
    \centering
    \includegraphics[width=0.2\textwidth]{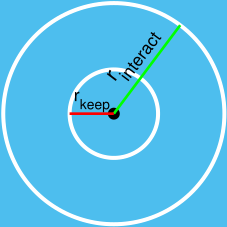}
  \caption{Regions of interaction, $r_{interact}$, and repulsion, $r_{keep}$, are centered on reference fish, $rf$ (black dot). \newline}
  \label{fig:regions}
\end{figure}

\begin{figure}[!ht]
    \begin{subfigure}[b]{0.45\textwidth}
 \centering   
    \includegraphics[width=0.5\textwidth]{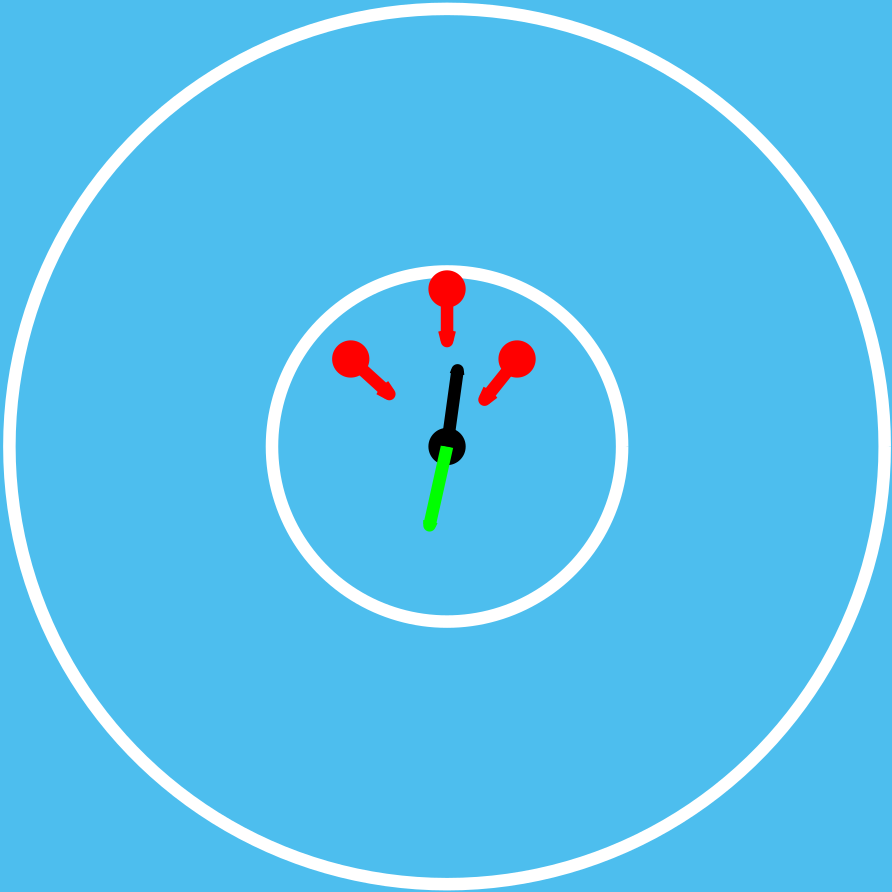}
  \caption{Directional change (from black to green arrow) to avoid too close schoolmates (red dots).\newline}
    \label{fig:collision}
  \end{subfigure}
  \begin{subfigure}[b]{0.45\textwidth}
   \centering
    \includegraphics[width=0.5\textwidth]{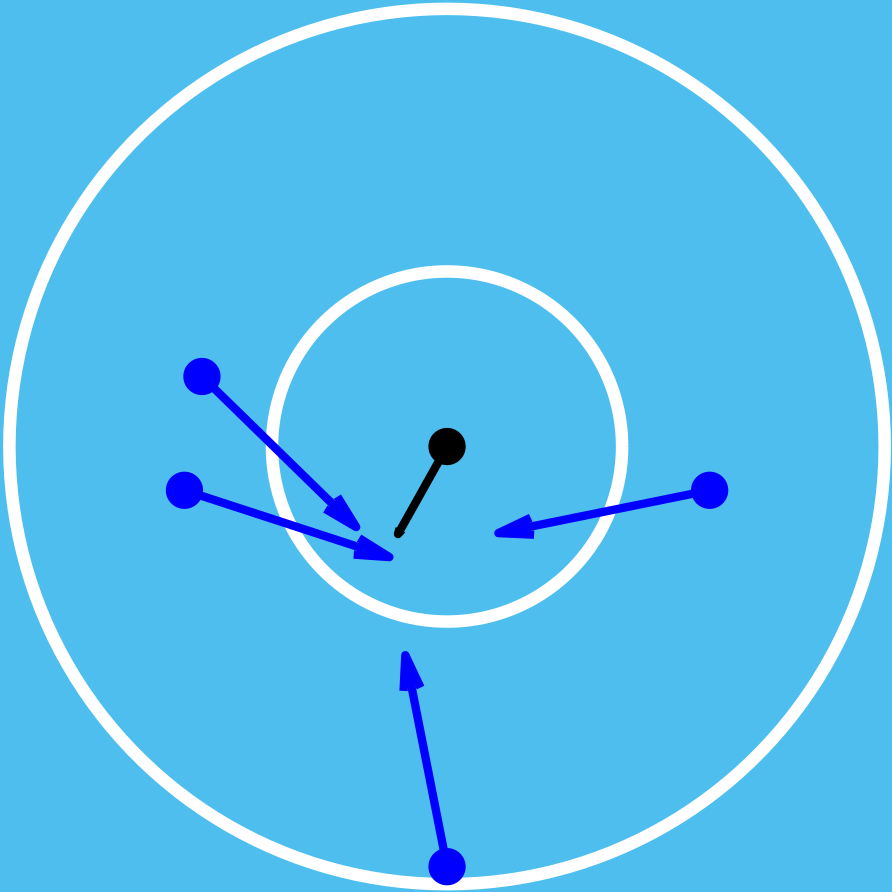}
  \caption{Movement towards the centroid of positions where schoolmates are located. \newline}
  \label{fig:cenering}
  \end{subfigure}
    \hfill
    \begin{subfigure}[b]{0.45\textwidth}
    \centering
    \includegraphics[width=0.5\textwidth]{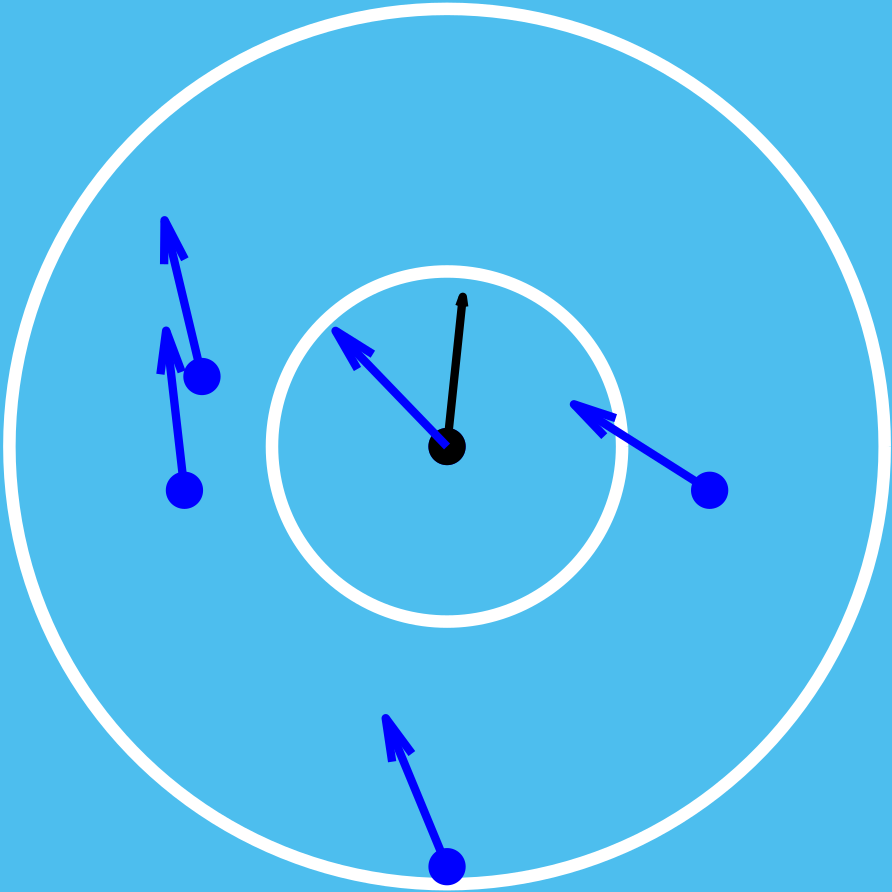}
  \caption{Directional change (from black to blue arrow) to align movement with schoolmates.}
  \label{fig:attraction}
  \end{subfigure}
    \label{fig:interaction}
    \caption{Fish local interaction rules -- the reference individual fish $rf$  (black dot) whose original direction (black arrow) is influenced by neighboring fish (red and blue dots) whose movement directions are indicated by red and blue arrows.}
\end{figure}

We follow the original Boid model \cite{reynolds1987flocks}, in which fish adhere to several interacting rules within a local range (see Fig.~\ref{fig:regions}). There is a cohesion rule that consists of two parts: (i) Collision avoidance, which represents a repulsive force ensuring a minimum degree of separation between schoolmates (see Fig.~\ref{fig:collision}). (ii) School attraction, which exerts a positive force to maintain the school's formation within the region of interaction (see Fig.~\ref{fig:attraction}). The combination of (i) and (ii) guarantees that the school formation occurs within a mutual range. The cohesion force vector is denoted as $F_{u_{c}}$ and is defined in Eq. (\ref{eq:cohs}).
\begin{equation}
F_{u_{c}}= \frac{u_{c}}{\left\Vert u_{c}\right\Vert}\cdot\begin{cases} \frac{\left\Vert u_{c}\right\Vert - \text{r}_{\mathrm{keep}}}{\text{r}_{\mathrm{keep}}} &\text{if}\ \left\Vert u_{c}\right\Vert\leq \text{r}_{\mathrm{interact}},\\  0  &\text{Elsewhere}.\\  \end{cases}
\label{eq:cohs}
\end{equation}
 where $u_c$ is the vector from the reference individual to the nearest schoolmate, $\text{r}_{\mathrm{keep}}$ is the minimum safe distance to avoid collision with a schoolmate, and $\text{r}_{\mathrm{interact}}$ is the maximum distance to interact (range of the region of interaction).
The alignment rule, which describes the propensity of an individual fish to move in alignment with the swimming velocity of its schoolmates, is defined by the force vector $F_{u_{a}}$, as given in Eq.(\ref{eq:ali}).
\begin{equation}
   F_{u_{a}} =  \frac 1 n \sum_{i \in R_n}  u_i -  u_{rf},
   \label{eq:ali}
\end{equation}
 where $u_i$ and $u_{rf}$ are the velocity vectors of $i$ and reference fish $rf$, respectively, and $R_n$ is the set of $n$ individuals located within the interaction region $rf$.
 
In addition to the foregoing two rules (school cohesion and alignment), which act in reaction to the environment, we can identify an attraction to stimuli (i.e., a positive force toward e.g food), and repulsion to risk sources (e.g predator) \cite{chen2006genetic}. The force vectors defining stimuli (attraction), $F_{u_{f}}$, and risk (repulsion), $F_{u_{r}}$, are given by Eq.(\ref{eq:food}) and Eq. (\ref{eq:risk}), respectively, where $u_f$ is the vector from the reference individual to the nearest stimuli sources, and $u_r$ is the vector from the reference individual to the nearest risk source.
 \begin{eqnarray}
 \label{eq:food}
   F_{u_{f}} &=&  \frac{u_{f}}{\left\Vert u_{f}\right\Vert},\\
\label{eq:risk}     
F_{u_{r}} &=& \frac{u_{r}}{\left\Vert u_{r}\right\Vert}\cdot\begin{cases} \frac{\left\Vert u_{r}\right\Vert - \text{r}_{\mathrm{interact}}}{\text{r}_{\mathrm{interact}}} &\text{if}\ \left\Vert u_{r}\right\Vert\leq \text{r}_{\mathrm{interact}},\\  0  &\text{Elsewhere}.\\  \end{cases}
\end{eqnarray}

The ultimate movement vector $V_{rf}$ of the $rf$ individual fish is defined as a weighted combination of all the influencing forces, which is expressed by Eq.~\ref{eq:finalF}, 
\begin{eqnarray}
\label{eq:finalF}
V_{rf}&=&\omega_{1}F_{u_{c}}+\omega_{2}F_{u_{a}}+\omega_{3} F_{u_{f}} + +\omega_{4} F_{u_{r}} 
\end{eqnarray}
where $\left\{\omega_{i}\right\}_{i=1}^4$, are the controlling weights that based on the spatiotemporal state of the individual surroundings. 

\section{K-means clustering and Voronoi tessellations}
\label{sec3} 
\begin{figure}[htbp]
    \centering
    \begin{tabular}{ccc}
        \includegraphics[width=0.22\textwidth]{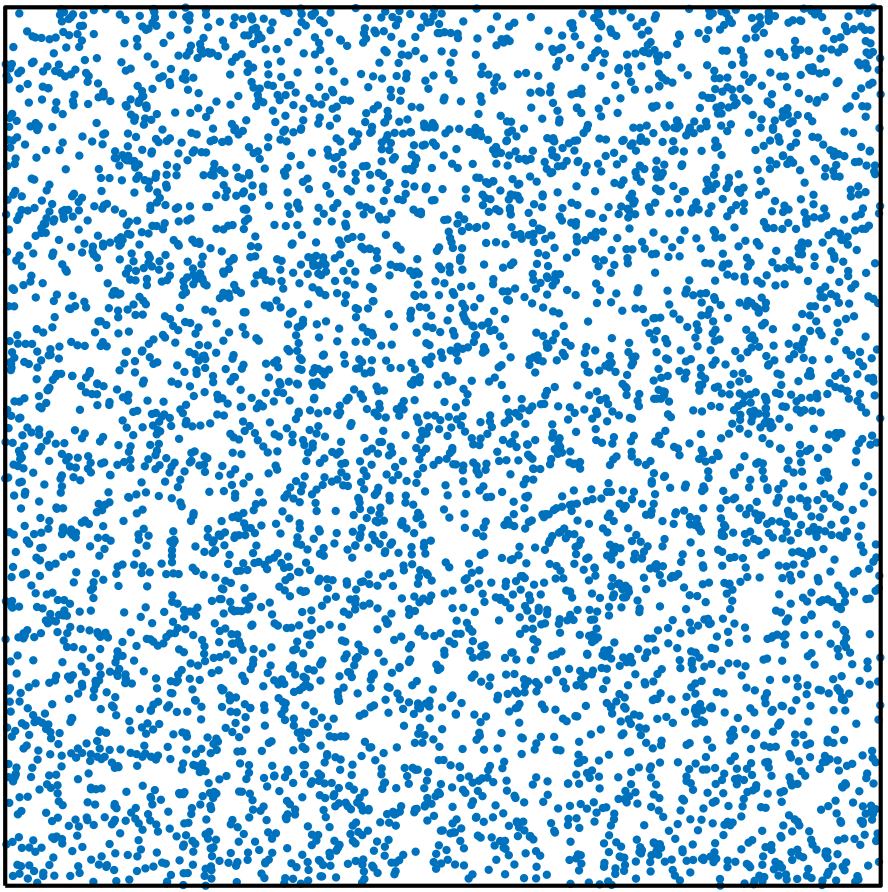} & $\vpointer$ & \includegraphics[width=0.2\textwidth]{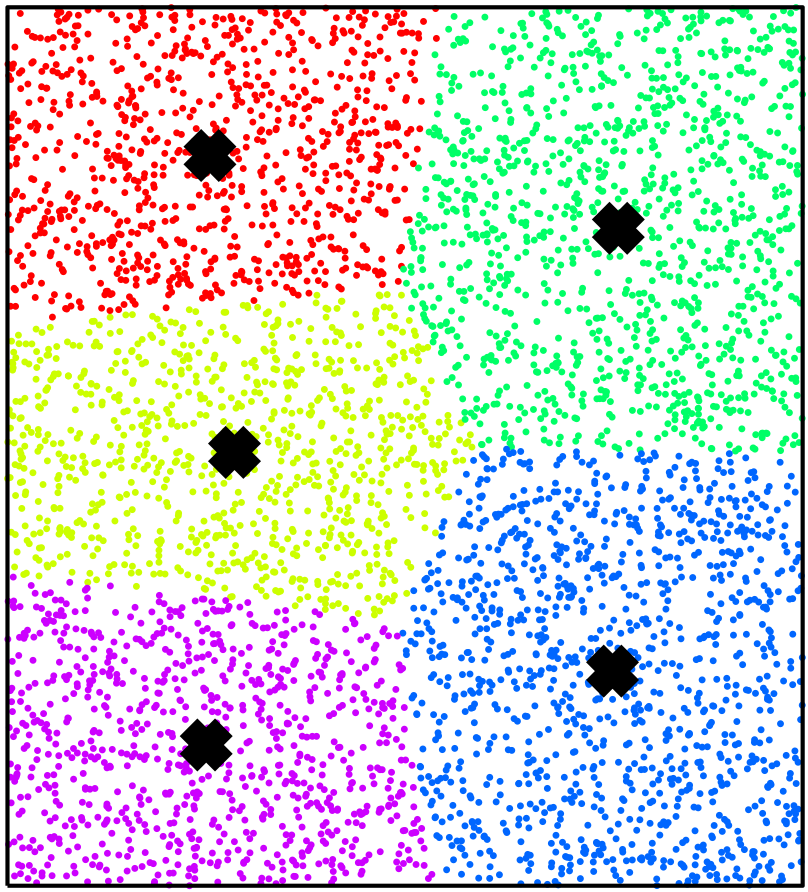} \\
        Raw data & & Clustered data
    \end{tabular}
    \caption{Illustration of K-means clustering}
    \label{fig:Pffft}
\end{figure}

\begin{figure}[htbp]
    \centering
    \begin{tabular}{ccc}
        \includegraphics[width=0.2185\textwidth]{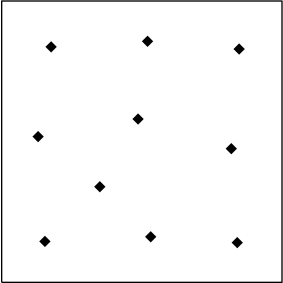} & $\vpointer$ & \includegraphics[width=0.2185\textwidth]{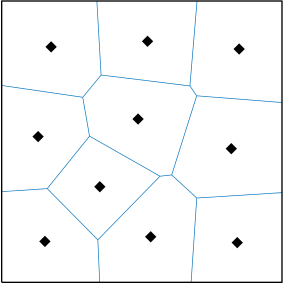} \\
        Random points & & Voronoi diagram
    \end{tabular}
    \caption{Illustration of Voronoi tessellation}
    \label{fig:voronoi}
\end{figure}

\begin{figure}[htbp]
    \centering
    \begin{tabular}{ccc}
        \includegraphics[width=0.215\textwidth]{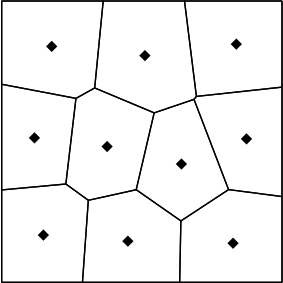} & $\vpointer$ & \includegraphics[width=0.215\textwidth]{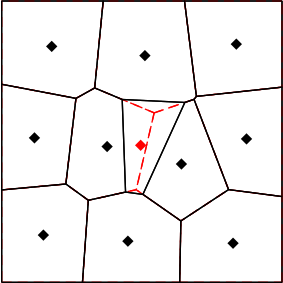} \\
        (a) VD -- 10 points & & (b) VD update -- 11 points
    \end{tabular}
    \caption{Update of Voronoi diagram (VD) after adding an additional point (red dot) to the diagam in (a).  The broken red lines in (b) are the original boundaries of the cells before the new point is added.}
    \label{fig:voronoi_geo}
\end{figure}

K-means is an unsupervised and non-deterministic learning algorithm widely used for clustering large datasets. It is considered one of the simplest clustering algorithms \cite{sun2008clustering, na2010research}. 
The algorithm works by initially assigning each data sample to the nearest centroid of a predefined number of centroids ($k$) within the dataset. The next step involves calculating the correct centroid for each data partition group and reassigning the samples to their nearest new centroids. This iterative process continues until there is no change in the centroids of the clusters \cite{franti2018k, sinaga2020unsupervised}. Figure~\ref{fig:Pffft} illustrates an initial dataset and the resulting clusters.
A Voronoi Diagram (\textit{VD})\cite{boots1999spatial,pokojski2018voronoi} is defined as a set of Voronoi cells $ VD = \bigg \{ V(p_1),V(p_2),..., V(p_n) \bigg \}$, generated by the points $p_1,p_2,...,p_n$, where, $d$ is the Euclidean distance,\\
$V(p_i) = \bigg\{ x |\; d(x,p_i) \le d(x,p_j) \;\; \text{for} j\neq i \bigg \}.$  \\
Voronoi cells are distinct, space-filling regions, and the size (volume or area) of each cell is inversely related to the density of the generating points, as shown in Fig.~\ref{fig:voronoi} and \ref{fig:voronoi_geo}. These characteristics of Voronoi cells are particularly appealing for high-dimensional computations \cite{subbey2003strategy}.

\section{Optimization algorithms}
\label{sec4}
\subsection{Evolutionary adaptation} 
We present the CPR algorithm, which combines the Voronoi tessellation (TS) and K-means clustering algorithms to quantify parameters (weights) associated with fish movement vectors. Algorithm~\ref{alg:cap} provides a skeletal representation of our algorithm. 

We partially adopt the principles of evolutionary algorithms when selecting the best candidate solutions. The genetic principle of mutation is applied when searching for potentially better candidates, which defines the new generation of solutions.

The algorithm is initiated by generating $N$ random numbers, $x$, in the search space $[0,1)^d$, where $d$ is the dimension of the search space. These numbers are then grouped into a predefined number of clusters, $k$ (where $k \leq N$), and the centroids $c_k$ are considered as candidate solutions for optimization. Next, we use Voronoi tessellation to divide the search space into $k$ sub-regions, denoted as $R_k$. The suitability of each sub-region is evaluated using a predefined performance assessment function (PAF), as discussed in the next subsection. For the next generation, we apply the following three steps to each of the $m$ chosen regions: generate $N$ random numbers, create $k$ clusters, generate $k$ sub-regions using Voronoi tessellation, and determine new candidates using the PAF.

In an approach similar to the Genetic Algorithm (GA) mutation process, we add random sub-regions from the worst (k-m) sub-regions to the search space at each generation. Unlike traditional GA optimization, our approach reduces the search space domain at each generation, facilitating faster convergence towards the optimal solution and its surroundings (the optimal region).
\begin{figure}[!ht]
\centering
\includegraphics[scale=0.35]{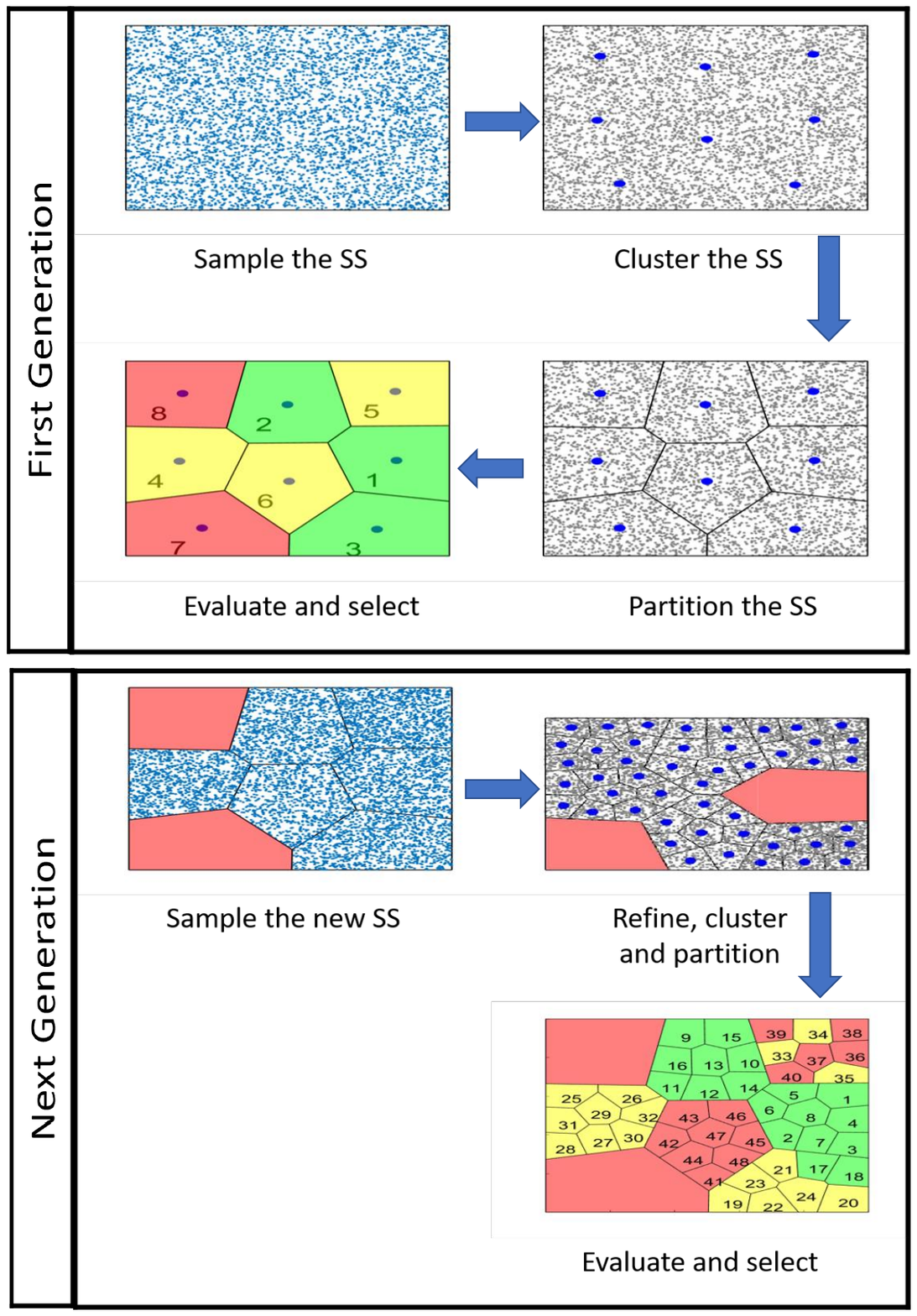}
   \caption{Schematic representation of the CPR algorithm: For each generation, random samples are generated from the search space (SS) and clustered (centroids defined by blue dots). The SS is then partitioned into bounded regions (black lines) using Voronoi tessellation. Finally, each region of the SS is evaluated based on the $PAF$ of its centroid. Based on the evaluation, regions are either removed from the search space (light red polygons) or promoted to become part of the new SS (yellow and light green polygons). Regions with a high $PAF$ (green polygons) are accepted directly, while regions with low $PAF$ (yellow polygons) may be added to the SS through random selection.} 
    \label{fig:my_label}
\end{figure}

\begin{algorithm}
\caption{Pseudocode of CPR algorithem}\label{alg:cap}
\begin{algorithmic}[1]
\STATE\textbf{Initialization}
 
$\textit{NumCand} = 20$;

k = \textit{NumCand};


$D_1= [0,1)^d$

\textit{NumSubRegions} = 1; 
  
\WHILE{max generations or optimal solution is found}
  
\STATE\textbf{Generate the solution candidates $c_k$:}

\STATE \textbf{ParFor} { i $\gets$ 1,NumSubRegions} 
\begin{itemize}
   \item[]  \textbf{Generate:} \\
    \textmd{ N (large enough) number of random points} $ x \sim \textit{Uniform}({D_i})$ 
   \item[] \textbf{Cluster:}\\
    \textmd{Find k number of clusters with centroids $c_k$
    (solution candidates) using the k-means clustering algorithm.}
     \item[] \textbf{Partition:} \\
   \textmd{ Construct k sub-regions $R_k$ based on VT method.} 
  \end{itemize}

\STATE\textbf{{Evaluate Candidates:}} 
    
    \textbf{ParFor} i $\gets$ 1, \textit{NumCand}
    
     \hspace{1cm} $PAF(i) = \textbf{SimulateFishSchool}(c_i)$\\
    \textbf{end for}
               
\STATE\textbf{Selection:}

select the best m candidates $c_m$, and the \\
\hspace{0.4cm}  corresponding m regions $R_m$.

\STATE\textbf{Mutation}:

Replace randomly p number of regions \\
(and their corresponding candidates $c_p$) 
from the best $ R_m $ by  p number of regions\\ 
 from the worst (\textit{NumCand} - m) regions. \newline
\STATE\textbf{Update parameters}

\textit{NumSubRegions} = k;

$D_i = R_i, \hspace{2dd} \forall i =1,2,...,k$.\\
k = 4
\ENDWHILE {}
\end{algorithmic}
\end{algorithm}

\subsection{Performance Assessment Function (PAF)}
\label{lab:subsec:PAF}
An optimal solution is one in which the derived (optimal) parameters result in schooling behavior that closely resembles empirical observations. However, when applying heuristic algorithms to quantify fish behavior, defining a realistic fitness function remains a challenge in behavioral ecology \cite{calvez2005automatic}. This paper adopts a fitness definition based on the collective fitness of the schooling fish, measured in terms of schooling coherence and proximity to the destination (e.g., a food source). Therefore, the PAF, as defined in (\ref{eq:paftxt}), represents the sum of individual fitness measures for each fish within the simulation population.
\begin{eqnarray}
    \label{eq:paftxt}
    PAF = \sum_{\forall Fish_i \in Population} \mathcal{F}( Fish_i),
\end{eqnarray}
where, for every individual fish, the function $\mathcal{F}$ consists of four components: 
\begin{eqnarray}
\label{eq:paf}
    \mathcal{F} =  f_1 + f_2 +f_3 +f_4
\end{eqnarray}

The first two components, $f_1$ and $f_2$, are related to collective assessment -- the degree of success in school formation. $f_1$ is computed in (Eq. \ref{eq:cost1}) by minimizing the divergence (angle) between the swimming velocity of the individual fish $ u_{rf}$ and the average velocity of schoolmates $u_{avg}$.
\begin{equation}
   f_1 =  \cos^{-1}\bigg(\frac{u_{avg} \cdot u_{rf}}{\Vert u_{avg} \Vert \times \Vert u_{rf} \Vert } \bigg ) 
   \label{eq:cost1}
\end{equation}
To assess the cohesion of the school and keeping it neither overcrowded nor scattered, we define the second component of the performance assessment function as in (Eq.\ref{eq:cost2}) 
\begin{equation}
f_2= \begin{cases}  \text{r}_{\mathrm{interact}} \bigg (\frac{ \Vert u_{c} \Vert  - \text{r}_{\mathrm{keep}}}{\text{r}_{\mathrm{keep}}}\bigg)^2 &\text{if}\ \left\Vert u_{c}\right\Vert\leq \text{r}_{\mathrm{interact}}\\
  \text{r}_{\mathrm{interact}}\bigg (\frac{ \Vert u_{c}\Vert  - \text{r}_{\mathrm{keep}}}{ \text{r}_{\mathrm{interact}} - \text{r}_{\mathrm{keep}}}\bigg)^2 &\text{if}\ \left\Vert u_{c}\right\Vert > \text{r}_{\mathrm{interact}} \\  \end{cases}
\label{eq:cost2}
\end{equation}

The third and fourth components, $f_3$ and $f_4$, assess the degree to which an individual fish successfully accomplishes its task, i.e., reaching the destination. The function in (Eq. \ref{eq:cost3}) minimizes the distance to the range, $[\text{r}_{\mathrm{f_1}},\text{r}_{\mathrm{f_2}}]$, of the attracting source (e.g food),  where $\text{r}_{\mathrm{f_1}} \le \text{r}_{\mathrm{f_2}} \le \text{r}_{\mathrm{interact}}$. 
\begin{equation}f_3= \begin{cases}  \text{r}_{\mathrm{interact}}  \bigg (\frac{\left\Vert u_{f}\right\Vert} {\text{r}_{\mathrm{f_1}}} - 1 \bigg )^2 &\text{if}\ \left\Vert u_{f}\right\Vert\leq \text{r}_{\mathrm{f_1}}\\  0  &\text{if}\  \text{r}_{\mathrm{f_1}} < \left\Vert u_{f}\right\Vert< \text{r}_{\mathrm{f_2}}, \\
\text{r}_{\mathrm{interact}}  \bigg (\frac{\left\Vert u_{f}\right\Vert} {\text{r}_{\mathrm{f_2}}} - 1 \bigg )^2 &\text{if}\ \left\Vert u_{r}\right\Vert\geq \text{r}_{\mathrm{f_2}}\\
\end{cases}
\label{eq:cost3}
\end{equation}
The fourth component, $f_4$, is related to the individual fish risk. The function in (Eq. \ref{eq:cost4}) is to minimize the risk of being preyed on, where ${\text{r}_{\mathrm{risk}}}$ is the maximum risk distance, and is bounded as ${\text{r}_{\mathrm{keep}}} \le {\text{r}_{\mathrm{risk}}} \le {\text{r}_{\mathrm{interact}}}$. 
\begin{equation}
f_4= \begin{cases}  \bigg (\frac{\left\Vert u_{r}\right\Vert} {\text{r}_{\mathrm{risk}}} - 1 \bigg )^2 &\text{if}\ \left\Vert u_{r}\right\Vert\leq \text{r}_{\mathrm{risk}}\\  0  &\text{Elsewhere}\\  \end{cases}
\label{eq:cost4}
\end{equation}
\section{Discussion of results from numerical experiments}
\label{sec5}
The primary goal of the numerical experiments was to demonstrate that parameters derived using the CPR algorithm result in schooling behavior that closely resembles empirical observations.

All algorithms and simulations were implemented on the MATLAB b2020b platform, running on a PC with an Intel Core i7, 2.6 GHz CPU, and 16 GB of RAM. To account for stochasticity, 20 independent runs were conducted for each experiment. In each simulation, the school consisted of 100 individual fish whose initial locations were randomly generated within the simulation domain. While, in general, $0 < r_{keep} < r_{interact}$, our specific choices of the radii, with $r_{interact}=3$ cm and $r_{keep}=1$ cm, were arbitrary.
\begin{figure}[!hb]
\centering
    \includegraphics[scale=0.65]{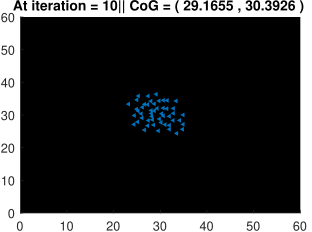}\hfill
    \includegraphics[scale=0.65]{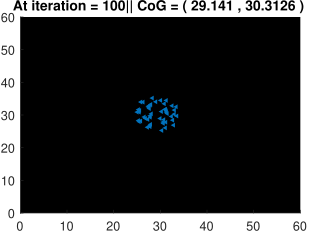}\hfill
    \includegraphics[scale=0.65]{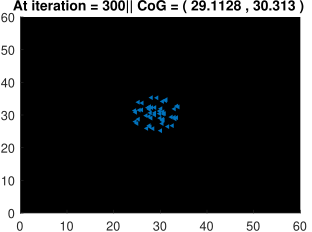}
    \caption{Schooling without optimized behavior rules}
     \label{fig:sim_NoVor}
\end{figure}

     \begin{figure}
     \centering
    \includegraphics[scale=0.65]{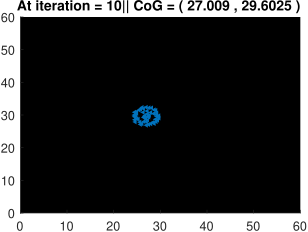}\hfill
    \includegraphics[scale=0.65]{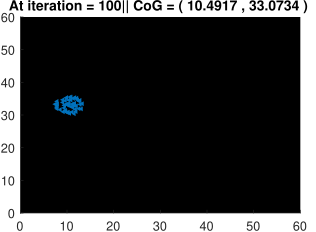}\hfill
    \includegraphics[scale=0.65]{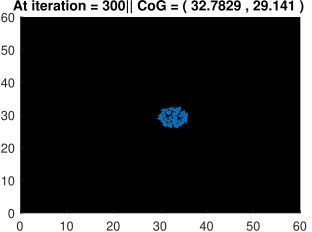}
    \caption{Schooling with optimized behavior rules}
    \label{fig:sim_Vor}
\end{figure}
The figures, Fig. \ref{fig:sim_NoVor} and Fig. \ref{fig:sim_Vor}, provide snapshots of the simulation results with and without CPR-based optimization. It is evident that without optimizing the rules, the school aggregates initially. However, the school becomes stagnant due to being caught between the attraction and repulsion forces (see Fig. \ref{fig:sim_NoVor}). On the other hand, using CPR-based optimization results in individual fish aggregating more quickly to form a larger group within the allowed distances, aligning their swimming directions and moving collectively (see Fig. \ref{fig:sim_Vor}). In other words, the schooling fish behavior is more realistic and closely resembles empirical observations. The density and size of the school of fish are determined by the predefined range of interaction, which varies between species.
\subsection{ CPR- vs GA- based optimization}
\begin{figure}[H]
    \centering
    \includegraphics[width=0.5\textwidth]{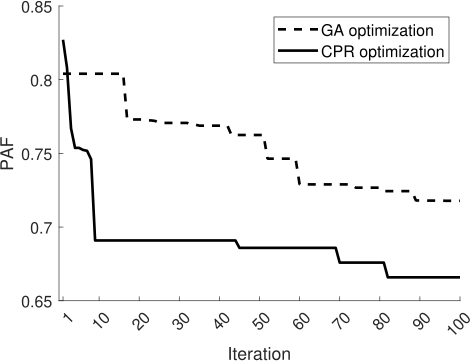}
    \caption{Comparison of performance profiles: CPR vs GA algorithms.}
\label{fig:performance}
\end{figure}

 We evaluated the performance of our algorithm by comparing its rate of convergence to an optimal solution with that of a Genetic Algorithm (GA). The parameters used for the comparison simulations are listed in Table~\ref{tab:para}.
\begin{table}[h!]
\centering
\caption{Optimization parameters}
\begin{tabular}{ c c c }
\hline
 Parameter & CPR  & GA \\ \hline
 Population size & 20 & 20 \\ 
 Search space & $ R_k \subseteq  [0,1)^d$, & $[0,1)^d$ \\
 Mutation rate  & 0.1 & 0.1 \\ 
 Selection rate  & 0.2 & 0.2 \\
\end{tabular}
\label{tab:para}
\end{table}  
The plot in Fig. \ref{fig:performance} displays the results obtained from 100 generations, with data averaged across ten simulation runs at each generation. The results clearly demonstrate that the CPR algorithm converges to the optimal solution more rapidly than conventional optimization using the GA. This accelerated convergence, as previously explained in a prior section, can be attributed to the narrowing of the search space as the algorithm evolves. Consequently, candidate solution spaces (Voronoi cells) achieve higher probabilities of containing the optimal solution at each generation.
\section{Conclusion}
In this paper, we introduced a CPU-efficient iterative search algorithm (CPR) for optimizing the decision rule parameters governing fish collective schooling behaviors. The CPR algorithm leverages unsupervised learning techniques and Voronoi tessellation to effectively partition and refine the search space. The algorithm was tested by simulating the collective schooling behavior of a large number of interacting particles. The results demonstrate that the schooling behavior generated by the CPR algorithm closely aligns with empirical observations. Furthermore, the CPR algorithm exhibits faster convergence compared to a standard Genetic Algorithm. Overall, the CPR algorithm presents a promising approach for efficiently simulating large-scale particle interactions.
\printbibliography

@inproceedings{reynolds1987flocks,
  title={Flocks, herds and schools: A distributed behavioral model},
  author={Reynolds, Craig W},
  booktitle={Proceedings of the 14th annual conference on Computer graphics and interactive techniques},
  pages={25--34},
  year={1987}
}

@inproceedings{chen2006genetic,
  title={Genetic algorithms for optimization of boids model},
  author={Chen, Yen-Wei and Kobayashi, Kanami and Huang, Xinyin and Nakao, Zensho},
  booktitle={International Conference on Knowledge-Based and Intelligent Information and Engineering Systems},
  pages={55--62},
  year={2006},
  organization={Springer}
}

@inproceedings{alaliyat2014optimisation,
  title={Optimisation Of Boids Swarm Model Based On Genetic Algorithm And Particle Swarm Optimisation Algorithm (Comparative Study).},
  author={Alaliyat, Saleh and Yndestad, Harald and Sanfilippo, Filippo},
  booktitle={ECMS},
  pages={643--650},
  year={2014},
  organization={Citeseer}
}

@misc{barksten2013extending,
  title={Extending Reynolds’ flocking model to asimulation of sheep in the presence of a predator},
  author={Barksten, Martin and Rydberg, David},
  year={2013}
}

@misc{chang2019investigating,
  title={Investigating and Modeling the Emergent Flocking Behaviour of Sheep Under Threat with Fear Contagion},
  author={Chang, Gabriel and Stjerndal, Michaela},
  year={2019}
}

@article{hartman2006autonomous,
  title={Autonomous boids},
  author={Hartman, Christopher and Benes, Bedrich},
  journal={Computer Animation and Virtual Worlds},
  volume={17},
  number={3-4},
  pages={199--206},
  year={2006},
  publisher={Wiley Online Library}
}

@article{tran2020switching,
  title={Switching formation strategy with the directed dynamic topology for collision avoidance of a multi-robot system in uncertain environments},
  author={Tran, Vu Phi and Garratt, Matthew A and Petersen, Ian R},
  journal={IET Control Theory \& Applications},
  volume={14},
  number={18},
  pages={2948--2959},
  year={2020},
  publisher={IET}
}

@article{erra2004massive,
  title={Massive simulation using gpu of a distributed behavioral model of a flock with obstacle avoidance},
  author={Erra, Ugo and De Chiara, Rosario and Scarano, Vittorio and Tatafiore, Maurizio},
  journal={Proceedings of Vision, Modeling and Visualization 2004 (VMV)},
  year={2004}
}

@article{goldberg1989genetic,
  title={Genetic algorithms in search, optimization, and machine learning. Addison},
  author={Goldberg, David E},
  journal={Reading},
  year={1989}
}

@article{forrest1993genetic,
  title={Genetic algorithms: principles of natural selection applied to computation},
  author={Forrest, Stephanie},
  journal={Science},
  volume={261},
  number={5123},
  pages={872--878},
  year={1993},
  publisher={American Association for the Advancement of Science}
}

@inproceedings{dmytruk2021safe,
  title={Safe Tightly-Constrained UAV Swarming in GNSS-denied Environments},
  author={Dmytruk, Andriy and Nascimento, Tiago and Ahmad, Afzal and Bavca, Tomavs and Saska, Martin},
  booktitle={2021 International Conference on Unmanned Aircraft Systems (ICUAS)},
  pages={1391--1399},
  year={2021},
  organization={IEEE}
}

@inproceedings{calvez2005automatic,
  title={Automatic tuning of agent-based models using genetic algorithms},
  author={Calvez, Beno{\^{i}}t and Hutzler, Guillaume},
  booktitle={International Workshop on Multi-Agent Systems and Agent-Based Simulation},
  pages={41--57},
  year={2005},
  organization={Springer}
}

@article{alaliyat2019optimal,
  title={Optimal fish densities and farm locations in Norwegian fjords: a framework to use a PSO algorithm to optimize an agent-based model to simulate fish disease dynamics},
  author={Alaliyat, Saleh and Yndestad, Harald and Davidsen, P{\aa}l I},
  journal={Aquaculture International},
  volume={27},
  number={3},
  pages={747--770},
  year={2019},
  publisher={Springer}
}

@article{samarasinghe2022grammar,
  title={Grammar-based cooperative learning for evolving collective behaviours in multi-agent systems},
  author={Samarasinghe, Dilini and Barlow, Michael and Lakshika, Erandi and Kasmarik, Kathryn},
  journal={Swarm and Evolutionary Computation},
  volume={69},
  pages={101017},
  year={2022},
  publisher={Elsevier}
}

@inproceedings{subbey2003strategy,
  title={A strategy for rapid quantification of uncertainty in reservoir performance prediction},
  author={Subbey, Sam and Mike, Christie and Sambridge, Malcolm},
  booktitle={SPE Reservoir Simulation Symposium},
  year={2003},
  organization={OnePetro}
}

@inproceedings{macqueen1967some,
  title={Some methods for classification and analysis of multivariate observations},
  author={MacQueen, James and others},
  booktitle={Proceedings of the fifth Berkeley symposium on mathematical statistics and probability},
  volume={1},
  number={14},
  pages={281--297},
  year={1967},
  organization={Oakland, CA, USA}
}

@book{boots2009spatial,
  title={Spatial tessellations: concepts and applications of Voronoi diagrams},
  author={Boots, Barry and Sugihara, Kokichi and Chiu, Sung Nok and Okabe, Atsuyuki},
  year={2009},
  publisher={John Wiley \& Sons}
}

@inproceedings{na2010research,
  title={Research on k-means clustering algorithm: An improved k-means clustering algorithm},
  author={Na, Shi and Xumin, Liu and Yong, Guan},
  booktitle={2010 Third International Symposium on intelligent information technology and security informatics},
  pages={63--67},
  year={2010},
  organization={Ieee}
}

@article{sun2008clustering,
  title={Clustering algorithms research},
  author={Sun, Ji-Gui and Liu, Jie and Zhao, Lian-Yu},
  journal={Journal of software},
  volume={19},
  number={1},
  pages={48--61},
  year={2008}
}

@article{franti2018k, 
  title={K-means properties on six clustering benchmark datasets},
  author={Fr{\"a}nti, Pasi and Sieranoja, Sami},
  journal={Applied Intelligence},
  volume={48},
  number={12},
  pages={4743--4759},
  year={2018},
  publisher={Springer}
}

@article{sinaga2020unsupervised,
  title={Unsupervised K-means clustering algorithm},
  author={Sinaga, Kristina P and Yang, Miin-Shen},
  journal={IEEE access},
  volume={8},
  pages={80716--80727},
  year={2020},
  publisher={IEEE}
}

@article{pokojski2018voronoi,
  title={Voronoi diagrams--inventor, method, applications},
  author={Pokojski, Wojciech and Pokojska, Paulina},
  journal={Polish Cartographical Review},
  volume={50},
  number={3},
  pages={141--150},
  year={2018}
}

@article{boots1999spatial,
  title={Spatial tessellations},
  author={Boots, B and Okabe, A and Sugihara, K},
  journal={Geographical information systems},
  volume={1},
  pages={503--526},
  year={1999},
  publisher={John Wiley \& Sons New York, NY}
}
\end{document}